\documentclass[12pt]{article}
\usepackage{graphicx}
\usepackage{amsmath,amsthm,amsfonts,amssymb}
\usepackage{color}
\usepackage[normalem]{ulem}
\makeatletter
\@addtoreset{equation}{section}

\makeatother

\begin{document}
\begin{titlepage}
\null
\begin{flushright}
\end{flushright}

\vskip 1.5cm
\begin{center}

 {\Large \bf Soliton Solutions of Noncommutative}

\vskip 0.4cm

 {\Large \bf  Anti-Self-Dual Yang-Mills Equations}

\vskip 1.5cm
\normalsize

{\large 
Claire R. Gilson$^\dagger$, 
Masashi Hamanaka$^{*}$,
\vskip 0.2cm
Shan-Chi Huang$^{*}$ 
and 
Jonathan J. C. Nimmo$^\dagger$\footnote{
Jonathan Nimmo sadly passed away on 20 June 2017. 
We owe section 4 to his unpublished note. 
}
 }

\vskip 10mm

        {\it $^*$Department of Mathematics, University of
 Nagoya,\\
                      Nagoya, 464-8602, JAPAN}

\vskip 0.5cm

        {\it $^\dagger$Department of Mathematics, University of
 Glasgow\\
              Glasgow G12 8QW, UK}

\vskip 1.5cm


{\bf \large Abstract}

\vskip 0.5cm
 
\end{center}

We present exact soliton solutions of anti-self-dual Yang-Mills 
equations for $G=GL(N)$ on noncommutative Euclidean spaces in four-dimension
by using the Darboux transformations. 
Generated solutions are represented by quasideterminants 
of Wronski matrices in compact forms.
We give special one-soliton solutions for $G=GL(2)$ 
whose energy density can be real-valued. 
We find that the soliton solutions are the same as the 
commutative ones and can be interpreted as 
one-domain walls in four-dimension. 
Scattering processes of the multi-soliton solutions are 
also discussed.


\end{titlepage}

\clearpage
\baselineskip 6.1mm


\section{Introduction}


Noncommutative extension of integrable systems 
has been studied for many years
and various integrability aspects have been revealed.
(See, e.g. \cite{CLPS, DiMH, DoLe, EGR, 
GHN, GHN2, KoOe, Kupershmidt, OlSo, ReRu} 
and references therein.)
In particular, exact soliton solutions have been found to be 
constructed in terms of quasideterminants in compact forms. 
The quasideterminants are introduced first by 
Gelfand and Retakh \cite{GeRe} 
toward a unification of theory of 
noncommutative determinants.
They actually play crucial roles in 
the theory of noncommutative solitons
and simplify proofs in commutative theories.
This suggests that quasideterminants might be 
fundamental objects which lead to a new essential
formulation of soliton theory. 

Darboux transformations are one of powerful 
methods to construct exact Wronskian-type solutions
of integrable systems \cite{MaSa}. 
This can be applied to
wide class of noncommutative integrable equations
in three-dimension of space-time or less. 
(See, e.g. \cite{GiMa, GiNi07, GNO07, GNS, HaHa, 
Hamanaka_JHEP, LNT, HSS} and references therein.)
For example, the noncommutative potential KP equation
is derived in the framework of the noncommutative
integrable hierarchy as follows:
\begin{eqnarray}
\label{ncKP}
\left(v_t+v_{xxx}+3v_x v_x \right)_x
 -3v_{yy}-3[v_x, v_y]=0,
\end{eqnarray}
where the subscripts denote partial derivatives. 
Every variable belongs to a noncommutative ring 
(e.g. quarternion).
Differentiation and intergration 
are assumed to be as in commutative case. 
The Wronskian solutions are first given \cite{EGR}
and derived by using the Darboux transformation 
in a natural way \cite{GiNi07}.

Extension of integrable equations to 
noncommutative (NC) spaces is also a hot topic.
(See, e.g. \cite{DiMH_KdV, Hamanaka_PhD, Hamanaka_PS, HaTo,  
Lechtenfeld, Lee, Ma, MaPa, Paniak, Sakakibara, Takasaki, WaWa} and references therein.)
In gauge theories, noncommutative extensions are 
equivalent to the presence of background $U(1)$ flux. 
In four dimension, anti-self-dual Yang-Mills (ASDYM)
equations are important in both mathematics and physics,
including integrable systems as well \cite{CFGY,MaWo,MOS,Takasaki_CMP,Ward}. 
Noncommutative anti-self-dual Yang-Mills equations 
have attracted a lot of interest in recent development
of gauge theories and string theories. 
By using noncommutative Penrose-Ward transformations, 
exact solutions are constructed in terms of 
quasideterminants which are not Wronskian-type \cite{GHN, GHN2}.
The solutions include not only instantons but 
soliton-type solutions. Two soliton scatterings 
are discussed for $G=GL(2)$ \cite{HaOk}, where 
energy densities (the second Chern classes) 
are in general complex valued. 
For $n$-soliton scatterings, Wronskian-type 
solutions are suitable. 
In order to interpret physical meaning of the solutions, 
energy densities should be real-valued.

In this paper, we construct exact multi soliton solutions
of anti-self-dual Yang-Mills equations for $G=GL(N)$
in four-dimensional noncommutative Euclidean spaces  
by generalizing the Darboux transformations 
\cite{GNO00} to noncommutative settings. 
The generated solutions are described in terms of quasideterminants
in compact forms. 
The proofs need only a few formula of quasideterminants. 
We focus on soliton solutions for $G=GL(2)$ 
where energy densities of them can be real.
We find that the one-soliton solutions have 
the same configurations as commutative ones which have 
localized energy densities on three-dimensional hyperplanes. 
These are domain-walls in $\mathbb{R}^4$. 
We show that asymptotic behaviors of multi-soliton solutions 
are the same as commutative ones. Therefore
the $n$ soliton solutions would be $n$ intersecting 
domain-walls with phase shifts. 
In discussion of the soliton scatterings, 
the properties of quasideterminants play a key role. 

This paper is organized as follows. 
In section 2, we introduce noncommutative anti-self-dual 
Yang-Mills equations in the star-product formalism. 
In section 3, we review definitions of quasi-determinants
and summarize some properties of them. 
In section 4, we present a noncommutative version 
of the Darboux transformation for $G=GL(N)$ 
noncommutative anti-self-dual 
Yang-Mills equations to generate exact solutions of them. 
We obtain the solutions of Yang's $J$ and $K$ matrices
in terms of quasideterminants. 
The results are all new and reduce to the commutative one \cite{GNO00}. 
In section 5, we give exact soliton solutions 
of the noncommutative anti-self-dual Yang-Mills equation for $G=GL(N)$
which are Wronskian-type.
We focus on $G=GL(2)$ case 
and present one soliton solutions 
with real-valued energy densities.
We also give multi soliton solutions of the same type 
and discuss the asymptotic behavior. 
The results are also brand new.
Section 6 is devoted conclusion and discussion.

\section{ASDYM Equations in Noncommutative Spaces}


Noncommutative spaces are defined
by the noncommutativity of the coordinates:
\begin{eqnarray}
\label{nc_coord}
[x^\mu,x^\nu]=i
\vartheta^{\mu\nu},
\end{eqnarray}
where the constant $\vartheta^{\mu\nu}$ is 
called the noncommutative parameter.
If the coordinates are real,
the noncommutative parameters should be real
because of hermicity of the coordinates.
We note that the noncommutative parameter 
$\vartheta^{\mu\nu}$ is
anti-symmetric with respect to $\mu$ and $\nu$ 
which implies that the rank of it is even.

Noncommutative field theories are given 
by the replacement of ordinary products
in the commutative field theories with the {star-products}.
The star-product is defined for ordinary c-number functions.
On flat spaces, is it represented explicitly by
\begin{eqnarray}
f\star g(x)&:=&\mbox{exp}
\left(\frac{i}{2}
\vartheta^{\mu\nu} \partial^{(x_1)}_\mu
\partial^{(x_2)}_\nu \right)
f(x_1)g(x_2)\Big{\vert}_{x_1=x_2=x}\nonumber\\
&=&f(x)g(x)+\frac{i}{2}
\vartheta^{\mu\nu}\partial_\mu f(x)\partial_\nu g(x)
+{\cal{O}} (\vartheta^2),
\label{star}
\end{eqnarray}
where $\partial_\mu^{(x)}:=\partial/\partial x^{\mu}$.
This is known as the {Moyal product} \cite{Moyal}.
The ordering of fields in nonlinear terms are determined
so that some structures such as gauge symmetries should be preserved.

The star-product has associativity:
$f\star(g\star h)=(f\star g)\star h$. 
It reduces to the ordinary product 
in the commutative limit:  $\vartheta^{\mu\nu}\rightarrow 0$.
In this sense, the noncommutative field theories
are deformed theories from the commutative ones.
The replacement of the product  makes the ordinary
spatial coordinates ``noncommutative,'' in the sense that
$[x^\mu,x^\nu]_\star:=x^\mu\star x^\nu-x^\nu\star x^\mu
=i\vartheta^{\mu\nu}$.

We note that the fields themselves take c-number values. 
The differentiation and the integration for them
are the same as commutative ones. 
Field equations are deformed and hence 
solutions of them are also deformed from commutative ones.


Now let us introduce the noncommutative 
anti-self-dual Yang-Mills equations in 
$4$-dimensional noncommutative Euclidean spaces 
whose real coordinates are denoted by $x^\mu~(\mu=1,2,3,4)$. 
The gauge group is denoted by $G$. 
In this paper we consider $G=GL(N,\mathbb{C})$ or subgroups of it.

Complex coordinates are introduced as follows:
\begin{eqnarray}
\label{cpx_coord}
y:=\frac{1}{\sqrt{2}}\left(x^1+ix^2\right),~
z:=\frac{1}{\sqrt{2}}\left(x^3+ix^4\right). 
\end{eqnarray}
The noncommutative anti-self-dual Yang-Mills equation is derived 
{}from the compatibility condition of the following linear system:
\begin{eqnarray}
L\star \phi&:=&(D_y-\zeta D_{\overline{z}})\star \phi=
 \left(\partial_y+A_y-\zeta (\partial_{\overline{z}}+A_{\overline{z}})\right)\star \phi(x;\zeta)
= 0,\nonumber\\
M\star \phi&:=&(D_z+\zeta D_{\overline{y}})\star \phi=
  \left(\partial_z+A_z+\zeta (\partial_{\overline{y}}+A_{\overline{y}})\right)\star \phi(x;\zeta)
= 0,
\label{lin_asdym}
\end{eqnarray}
where $A_z,A_y,A_{\overline{z}},A_{\overline{y}}$ 
and $D_z,D_y,D_{\overline{z}},D_{\overline{y}}$  
denote gauge fields and covariant derivatives in the Yang-Mills theory,
respectively. The constant $\zeta\in \mathbb{C}P_1$ is called
the {\it spectral parameter}. 

The compatibility condition $[L,M]_\star=0$, gives rise to
a quadratic polynomial of $\zeta$ and each coefficient
yields the anti-self-dual Yang-Mills equation:
\begin{eqnarray}
F^\star_{yz}=0,~~~
F^\star_{\overline{y}\,\overline{z}}=0,~~~
F^\star_{z\overline{z}}+F^\star_{y\overline{y}}=0,
\label{asdym}
\end{eqnarray}
where $F^\star_{\mu\nu}:=\partial_\mu A_\nu-\partial_\nu A_\mu+[A_\mu,A_\nu]_\star$ denotes the field strength. This is equivalent to 
self-duality in the sense of the Hodge dual operator $*$:
$F^\star_{\mu\nu}=-*F^\star_{\mu\nu}$. 


Gauge transformations act on the linear system (\ref{lin_asdym}) as
\begin{eqnarray}
 L\mapsto g^{-1}\star L\star g,~
 M\mapsto g^{-1}\star M\star g,~
 \phi\mapsto g^{-1}\star \phi,~~~g(x)\in G.
\end{eqnarray}

Here we discuss the potential forms of the 
noncommutative anti-self-dual Yang-Mills equations
such as noncommutative $J,K$-matrix formalism and 
the noncommutative Yang's equation \cite{Takasaki}.

The {\it $J$-matrix formalism} is given as follows.
The first equation of noncommutative anti-self-dual Yang-Mills equation
(\ref{asdym}) is the compatible condition
of the linear system $D_z\star h=0, D_y\star h=0$,
where $h$ is a $N\times N$ matrix.
Hence we get
$A_{z}=-(\partial_z h)\star h^{-1}, ~
A_y=-(\partial_y h)\star h^{-1}$.
Similarly, the second equation of noncommutative anti-self-dual 
Yang-Mills equation (\ref{asdym}) leads to
$
A_{\overline{z}}=-(\partial_{\overline{z}}\widetilde{h})\star \widetilde{h}^{-1}, ~
A_{\overline{y}}=(\partial_{\overline{y}}\widetilde{h})\star \widetilde{h}^{-1},
$ where $\widetilde{h}$ is also a $N\times N$ matrix.
We note that $h(x)=\phi(x,\zeta=0), \widetilde{h}(x)=\widetilde{\phi}(x,\zeta=\infty)$.

By defining a new matrix $J=\widetilde{h}\star h^{-1}$,
the third equation of the noncommutative anti-self-dual Yang-Mills equation 
(\ref{asdym}) becomes the noncommutative Yang's equation 
\begin{eqnarray}
\label{yang}
 \partial_{\overline{z}}(\partial_z J \star J^{-1})
+\partial_{\overline{y}} (\partial_y J \star J^{-1} )=0.
\end{eqnarray}

Gauge transformations act on $h$ and $\widetilde{h}$ as $h\mapsto g^{-1}\star 
h,~ \widetilde{h}\mapsto g^{-1}\star \widetilde{h},~g(x) \in G$.
Hence the Yang's $J$-matrix is gauge invariant. 
We note that all $h, \widetilde{h}$ and $J$ take values in $G$.


There is another potential form of the 
noncommutative anti-self-dual Yang-Mills equation,
known as the {\it $K$-matrix formalism}. 
In the gauge $A_{\overline{y}}=A_{\overline{z}}=0$, 
the third equation of (\ref{asdym}) becomes 
$\partial_{\overline{z}} A_z+\partial_{\overline{y}} A_y=0$.
This implies the existence of a potential $K$ such
that $A_z=-\partial_{\overline{y}}K,A_y=\partial_{\overline{z}}K$.
Then the second equation of (\ref{asdym}) becomes
\begin{eqnarray}
 \partial_y\partial_{\overline{y}}K +\partial_z\partial_{\overline{z}}K +[\partial_{\overline{z}} K, \partial_{\overline{y}} K]_\star=0.
\end{eqnarray}
Then, we get
\begin{eqnarray}
A_z = -\partial_zJ\star J^{-1} = -\partial_{\overline{y}} K,~~~
A_y= -\partial_y J\star J^{-1} = \partial_{\overline{z}} K.
\end{eqnarray}

In gauge theory, gauge invariant quantities are important.
In this paper, we will discuss the following quantity,
which is called the energy density,
\begin{eqnarray}
{\mbox{Tr}} F_\star^2 &:=& {\mbox{Tr}} F^\star_{\mu\nu}\star F_\star^{\mu\nu}.
\end{eqnarray}
We follow the Einstein convention. 
In the right hand side, the summation over $\mu,\nu=1,2,3,4$ is taken.


\section{Review of Quasi-determinants}

In this section, we give a quick review of 
quasi-determinants introduced by Gelfand and Retakh
\cite{GeRe} and present a few properties
of them which play key roles in the following sections.
Detailed discussion is seen in e.g. \cite{GGRW}.

Quasi-determinants are not just a generalization of
usual commutative determinants but rather
related to inverse matrices. 
{}From now on, we assume existence of the inverses in any case.

Let $A=(a_{ij})$ be a $N\times N$ matrix and 
$B=(b_{ij})$ be the inverse matrix of $A$: $A\star B=B\star A =1$.
In this paper, all products of matrix elements are assumed to be
star-products. 

Quasi-determinants of $A$ are defined formally
as the inverse of the elements of $B=A^{-1}$:
\begin{eqnarray}
 \vert A \vert^\star_{ij}:=b_{ji}^{-1}.
\end{eqnarray}
In the commutative limit, this is reduced to
\begin{eqnarray}
 \vert A \vert^\star_{ij} \stackrel{\theta\rightarrow 0}{\longrightarrow}
  (-1)^{i+j}\frac{\det A}{\det {A}^{ij}},
\label{limit}
\end{eqnarray}
where ${A}^{ij}$ is the matrix obtained from $A$
by deleting the $i$-th row and the $j$-th column.

We can write down a more explicit form of quasi-determinants.
In order to see it, let us recall the following formula
for the inverse $2\times 2$ block matrix:
\begin{eqnarray*}
 \left[
 \begin{array}{cc}
  A&B \\C&d
 \end{array}
 \right]^{-1}
=\left[\begin{array}{cc}
A^{-1}+A^{-1}\star B\star s^{-1}\star C \star A^{-1}
 &-A^{-1}\star B\star s^{-1}\\
 -s^{-1}\star C\star A^{-1}
&s^{-1}
\end{array}\right],
\end{eqnarray*}
where $A$ is a square matrix and $d$ is a single element and 
$s:=d-C\star A^{-1}\star B$ is called the Schur complement. 
We note that any matrix can be decomposed
as a $2\times 2$ matrix by block decomposition
where one of the diagonal parts is $1 \times 1$.
By choosing an appropriate partitioning,
any element in the inverse of a square matrix can be expressed
as the inverse of the Schur complement.
Hence quasi-determinants can be defined iteratively by:
\begin{eqnarray}
 \vert A \vert^\star_{ij}&=&a_{ij}-\sum_{i^\prime (\neq i), j^\prime (\neq j)}
  a_{ii^\prime} \star (({A}^{ij})^{-1})_{i^\prime j^\prime} \star
  a_{j^\prime
  j}\nonumber\\
 &=&a_{ij}-\sum_{i^\prime (\neq i), j^\prime (\neq j)}
  a_{ii^\prime} \star (\vert {A}^{ij}\vert_{j^\prime i^\prime })^{-1}
  \star a_{j^\prime j}.
\end{eqnarray}
It is convenient to represent the quasi-determinant
as follows:
\begin{eqnarray}
\label{Q-det}
 \vert A\vert^\star_{ij}=
\left|
  \begin{array}{ccccc}
   a_{11}&\cdots &a_{1j} & \cdots& a_{1n}\\
   \vdots & & \vdots & & \vdots\\
   a_{i1}&~ & {\fbox{$a_{ij}$}}& ~& a_{in}\\
   \vdots & & \vdots & & \vdots\\
   a_{n1}& \cdots & a_{nj}&\cdots & a_{nn}
  \end{array}\right|_\star.
\end{eqnarray}

Examples of quasi-determinants are,
for a $1\times 1$ matrix $A=a$
 \begin{eqnarray*}
  \vert A \vert^\star  = a,
 \end{eqnarray*}
and 
for a $2\times 2$ matrix $A=(a_{ij})$
 \begin{eqnarray*}
  \vert A \vert^\star_{11}=
   \begin{vmatrix}
   \fbox{$a_{11}$} &a_{12} \\a_{21}&a_{22}
   \end{vmatrix}_\star
 =a_{11}-a_{12}\star a_{22}^{-1}\star a_{21},~~~
  \vert A \vert^\star_{12}=
   \begin{vmatrix}
   a_{11} & \fbox{$a_{12}$} \\a_{21}&a_{22}
   \end{vmatrix}_\star
 =a_{12}-a_{11}\star a_{21}^{-1}\star a_{22},\nonumber\\
  \vert A \vert^\star_{21}=
   \begin{vmatrix}
   a_{11} &a_{12} \\ \fbox{$a_{21}$}&a_{22}
   \end{vmatrix}_\star
 =a_{21}-a_{22}\star a_{12}^{-1}\star a_{11},~~~
  \vert A \vert^\star_{22}=
   \begin{vmatrix}
   a_{11} & a_{12} \\a_{21}&\fbox{$a_{22}$}
   \end{vmatrix}_\star
 =a_{22}-a_{21}\star a_{11}^{-1}\star a_{12}, 
 \end{eqnarray*}
 and for a $3\times 3$ matrix $A=(a_{ij})$
  \begin{eqnarray*}
  \vert A \vert^\star_{11}
   &=&
   \begin{vmatrix}
   \fbox{$a_{11}$} &a_{12} &a_{13}\\ a_{21}&a_{22}&a_{23}\\a_{31}&a_{32}&a_{33}
   \end{vmatrix}_\star
=a_{11}-(a_{12}, a_{13})\star \left(
\begin{array}{cc}a_{22} & a_{23} \\a_{32}&a_{33}\end{array}\right)^{-1}
\star \left(
\begin{array}{c}a_{21} \\a_{31}\end{array}
\right)
\nonumber\\
  &=&a_{11}-a_{12}\star  \begin{vmatrix}
                   \fbox{$a_{22}$} & a_{23} \\a_{32}&a_{33}
                   \end{vmatrix}^{-1}_\star  \star a_{21}
           -a_{12}\star \begin{vmatrix}
                   a_{22} & a_{23} \\\fbox{$a_{32}$}&a_{33}
                   \end{vmatrix}^{-1}_\star \star a_{31}      \nonumber\\
&&~~~~    -a_{13}\star \begin{vmatrix}
                   a_{22} & \fbox{$a_{23}$} \\a_{32}&a_{33}
                  \end{vmatrix}_\star^{-1}\star  a_{21}
           -a_{13}\star \begin{vmatrix}
                   a_{22} & a_{23} \\a_{32}&\fbox{$a_{33}$}
                   \end{vmatrix}^{-1}_\star \star a_{31},
\end{eqnarray*}
and so on.

Quasideterminants have various interesting properties
similar to those of determinants. Among them, 
the following ones are relevant to 
the discussion on soliton scattering.

\vspace{2mm}
\noindent
{\bf Proposition 3.1 \cite{GeRe}}
Let $A=(a_{ij})$ be a square matrix of order $n$.

\noindent 
(i) {Permutation of Rows and Columns}.

The quasi-determinant $\vert A\vert^\star_{ij}$
does not depend on permutations of rows and columns
in the matrix $A$ that do not involve the $i$-th row
and $j$-th column.

\noindent
(ii) {The multiplication of rows and columns}.

Let the matrix $M=(m_{ij})$ be obtained from
the matrix $A$ by multiplying the $i$-th row by
$f(x)$ from the left, that is,
$m_{ij}=f \star a_{ij}$ and $m_{kj}=a_{kj}$
for $k\neq i$. Then
\begin{eqnarray}
 \vert M\vert^\star_{kj}=\left\{
                   \begin{array}{ll}
                    f \star\vert A \vert^\star_{ij}& \mbox{for}~ k=i 
                     \\ \vert A \vert^\star_{kj}& \mbox{for}~ k\neq i~~\mbox{and
                     }~f~ \mbox{is invertible}
                   \end{array}\right.
\end{eqnarray}

Let the matrix $N=(n_{ij})$ be obtained from
the matrix $A$ by multiplying the $j$-th column by
$f(x)$ from the right, that is,
$n_{ij}=a_{ij}\star f $ and $n_{il}=a_{il}$
for $l\neq j$. Then
\begin{eqnarray}
 \vert N\vert{^\star}_{il}=\left\{
                   \begin{array}{ll}
                    \vert A \vert^\star_{ij}\star f& \mbox{for}~ l=j
                     \\ \vert A \vert^\star_{il}& \mbox{for}~ l\neq j~~\mbox{and
                     }~ f ~\mbox{is invertible} 
                   \end{array}\right.
\end{eqnarray}

\noindent
(iii) {The addition of rows and columns}.

Let the matrix $M=(m_{ij})$ be obtained from
the matrix $A$ by replacing the $k$-th row of $A$
with the sum of the $k$-th row and $l$-th row, that is,
$m_{kj}=a_{kj}+a_{{lj}}$ and $m_{ij}=a_{ij}$
for $k\neq i$. Then
\vspace{-2mm}
\begin{eqnarray}
 \vert A\vert^\star_{ij}=\vert M \vert^\star_{ij},
  ~~~\mbox{for}~i\neq k.
\end{eqnarray}

Let the matrix $N=(n_{ij})$ be obtained from
the matrix $A$ by replacing the $k$-th column of $A$
with the sum of the $k$-th column and $l$-th column, that is,
$n_{ik}=a_{ik}+a_{il}$ and $n_{ij}=a_{ij}$
for $k\neq j$. Then
\vspace{-2mm}
\begin{eqnarray}
 \vert A\vert^\star_{ij}=\vert N \vert^\star_{ij},
  ~~~\mbox{for}~ j\neq k.
\end{eqnarray}

 \noindent 
{\bf Proposition 3.2 \cite{GeRe}} 
If the quasi-determinant $\vert A\vert_{ij}$
is defined, then the following statements are equivalent.

\noindent
(i) $\vert A \vert^\star_{ij}=0$.

\noindent
(ii) the $i$-th row of the matrix $A$ is a left linear combination
of the other rows of $A$.

\noindent
(iii) the $j$-th column of the matrix $A$
is a right linear combination of the other columns of $A$.

\vspace{2mm}
 \noindent 
{\bf Proposition 3.3} 
In the block matrices given in the following results, 
lower case letters denote single entries and 
upper case letters denote matrices of compatible dimensions 
so that the overall matrix is square. 

\vspace{1mm}
\noindent
(i) NC Jacobi identity \cite{GiNi07} 
(A useful special case of the NC Sylvester's Theorem\cite{GeRe}):
\begin{equation}\label{nc_syl}
    \begin{vmatrix}
      A&B&C\\
      D&f&g\\
      E&h&\fbox{$i$}
    \end{vmatrix}_\star=
    \begin{vmatrix}
      A&C\\
      E&\fbox{$i$}
    \end{vmatrix}_\star-
    \begin{vmatrix}
      A&B\\
      E&\fbox{$h$}
    \end{vmatrix}_\star
\star    
\begin{vmatrix}
      A&B\\
      D&\fbox{$f$}
    \end{vmatrix}_\star^{-1}
\star
    \begin{vmatrix}
      A&C\\
      D&\fbox{$g$}
    \end{vmatrix}_\star.
\end{equation}

\noindent
(ii) Homological relations \cite{GeRe}
\begin{eqnarray}
    \begin{vmatrix}
      A&B&C\\
      D&f&g\\
      E&\fbox{$h$}&i
    \end{vmatrix}_\star
\!\!\!=\!  \begin{vmatrix}
      A&B&C\\
      D&f&g\\
      E&h&\fbox{$i$}
    \end{vmatrix}_\star
\!\!\!\star\!
    \begin{vmatrix}
      A&B&C\\
      D&f&g\\
      0&\fbox{0}&1
    \end{vmatrix},~
\label{homological}
    \begin{vmatrix}
      A&B&C\\
      D&f&\fbox{$g$}\\
      E&h&i
    \end{vmatrix}_\star
\!\!\!=\!      \begin{vmatrix}
      A&B&0\\
      D&f&\fbox{0}\\
      E&h&1
    \end{vmatrix}_\star
\!\!\!\star\! 
    \begin{vmatrix}
      A&B&C\\
      D&f&g\\
      E&h&\fbox{$i$}
    \end{vmatrix}_\star
\end{eqnarray}

\noindent
(iii)  A derivative formula for quasideterminants \cite{GiNi07}
\begin{align}
    \begin{vmatrix}
    A&B\\
    C&\fbox{$d$}
    \end{vmatrix}^{\prime}_\star
 \label{row_diff} &=
    \begin{vmatrix}
    A&B\\
    C'&\fbox{$d'$}
    \end{vmatrix}_\star
    +\sum_{k=1}^n
    \begin{vmatrix}
    A&e_k\\
    C&\fbox{$0$}
    \end{vmatrix}_\star
\star
    \begin{vmatrix}
    A&B\\
    (A_k)'&\fbox{$(B_k)'$}
    \end{vmatrix}_\star,\\
 \label{col_diff} &=
    \begin{vmatrix}
    A&B'\\
    C&\fbox{$d'$}
    \end{vmatrix}_\star
    +\sum_{k=1}^n
    \begin{vmatrix}
    A&(A_k)'\\
    C&\fbox{$(C_k)'$}
    \end{vmatrix}_\star
\star
    \begin{vmatrix}
    A&B\\
    e_k^t&\fbox{$0$}
    \end{vmatrix}_\star,
\end{align}
where $A_k$ is the $k$th column and 
$A^k$ is the $k$th row of a matrix $A$ and 
$e_k$ is the column $n$-vector $(\delta_{ik})$ (i.e.\ 1 in the
$k$th row and 0 elsewhere). 


We note that the definition of the quasideterminants
are valid in the case that the elements of matrices belong to
noncommutative associative algebras, 
for example, the case that all elements $a_{ij}$ in \eqref{Q-det}
are $N\times N$ matrices. Propositions 3.1-3.3 still hold.  
In the next section, we will consider such situations.

\vspace{3mm}
\section{Darboux Transformation for NC ASDYM equation}

In this section, we give the Darboux transformation
for the $G=GL(N)$ noncommutative anti-self-dual Yang-Mills equation.

Let us start with the following linear systems:
\begin{eqnarray}
L\star \phi&:=&
J\star \partial_{y}(J^{-1}\star \phi)
- (\partial_{\overline{z}}\phi)\zeta=0,\nonumber\\
M\star \phi&:=&
J\star \partial_{z}(J^{-1}\star \phi)
+ (\partial_{\overline{y}}\phi)\zeta=0,
\label{lin_yang}
\end{eqnarray}
where $\zeta$ is 
a matrix generalization of the spectral parameter
and $N\times N$ constant matrix.
We note that this matrix acts on $\phi$ from right side
where $\phi$ is an $N\times N$ matrix 
whose columns consist of $N$-independent solutions of 
the linear systems. We can find that 
the compatibility condition 
$L\star M\star \phi -M\star L\star \phi=0$ 
gives the noncommutative anti-self-dual Yang-Mills equation.

Here the existence of $N$-independent solutions of 
the linear systems is assumed 
because of lack of Frobenius theorem. 
We will see in Theorem {5.1} that there exists 
$N\times N$ matrix $\phi$ which leads to soliton-type solutions
when the spectral parameter matrix $\zeta$ is diagonal.

\vspace{3mm}
\noindent
{\bf Theorem 4.1}
The linear system \eqref{lin_yang} is covariant under
the following Darboux transformation:
\begin{eqnarray}
\label{Darboux_phi}
 \widetilde{\phi}&=&
\phi \zeta - \theta \Lambda\star \theta^{-1}\star \phi,
\\
\label{Darboux_J}
\widetilde{J}&=& -\theta \Lambda\star \theta^{-1}\star J,
\end{eqnarray}
when $\theta$ is an eigenfunction of the linear system
\eqref{lin_yang}
for the choice of eigenvalue $\zeta=\Lambda$. 
We use the notation $\phi^{(k)}:=\phi \zeta^k$, where
$\phi$ and $\zeta$ are any eigenfunction-eigenvalue pair. 
(Hence $\theta^{(k)}:=\theta \Lambda^k$.)

\vspace{3mm}
\noindent
(Proof) It is proved by straightforward calculation
by using original conditions $
J\star\partial_y(J^{-1}\star \phi)-(\partial_{\overline{z}}\phi)\zeta=0$
and $J\star\partial_y(J^{-1}\star \theta)-(\partial_{\overline{z}}\theta) 
\Lambda=0$ and Eqs.(\ref{Darboux_phi}), (\ref{Darboux_J}): 
\vspace{3mm}
\begin{eqnarray*}
\widetilde{L} \widetilde{\phi}
\!&=&\! 
\widetilde{J}\star\partial_y
(\widetilde{J}^{-1}\!\star \widetilde{\phi})
\!-\!(\partial_{\overline{z}}\widetilde{\phi})\zeta\\
\!&=&\!
-\theta \Lambda\star \theta^{-1}\star J\star
\partial_y(-J^{-1}\!\star \theta \Lambda^{-1}\star \! \theta^{-1}\star \phi\zeta
+J^{-1}\!\star\phi)
\!-\!\partial_{\overline{z}}(\phi \zeta - 
\theta \Lambda\star \! \theta^{-1}\star \phi)\zeta\\
&=&
\theta\Lambda\star\theta^{-1}\star[J\star(\partial_yJ^{-1})\star\theta + \partial_y\theta]  
\Lambda^{-1}\star\theta^{-1}\star\phi\zeta + \theta\star(\partial_y\theta^{-1})\star\phi\zeta
+ \partial_y(\phi\zeta)
\\
& &-\theta\Lambda\star\theta^{-1}\star J\star\partial_y(J^{-1}\star\phi)
\\
& &
-(\partial_{\overline{z}}\phi)\zeta^2 + (\partial_{\overline{z}}\theta)\Lambda\star\theta^{-1}\star\phi\zeta
+ \theta\Lambda\star(\partial_{\overline{z}}\theta^{-1})\star\phi\zeta +
\theta\Lambda\star\theta^{-1}\star(\partial_{\overline{z}}\phi)\zeta
\\
&=&
\theta\Lambda\star \theta^{-1}\star
\left\{
-J\star \partial_y (J^{-1} \star \phi)
+(\partial_{\overline{z}} \phi)\zeta\right\} 
\\
&&
+\theta\Lambda\star \theta^{-1}\star
\left\{\partial_y \theta-(\partial_y J)\star J^{-1}
\star \theta-(\partial_{\overline{z}} \theta)\Lambda 
\right\} \Lambda^{-1}\star  \theta^{-1}\star \phi { \zeta}
\\
&& 
+\left\{-\partial_y\theta
        -J\star (\partial_y J^{-1})\star \theta
                  +(\partial_{\overline{z}}\theta) \Lambda 
\right\}
\star\theta^{-1}\star \phi \zeta=0
\end{eqnarray*}
In the {last} line, we substitute 
$\partial_{y}\phi -\partial_{\overline{z}}\phi \zeta$
with $-J\star (\partial_y J^{-1})\star\phi$
by using the original condition. 
It is similar to show $\widetilde{M} \widetilde{\phi}=0$. $\square$

Iterating the Darboux transformation \eqref{Darboux_phi}
and \eqref{Darboux_J}, we get solutions of the linear systems.

\vspace{3mm}
\noindent
{\bf Theorem 4.2}
Let us prepare eigenfunction-eigenvalue pairs $(\theta_i, \Lambda_i)~
i=1,2,\cdots, n$, and consider $n$ iteration of 
the Darboux transformation \eqref{Darboux_phi}:
\begin{eqnarray}
\label{Darboux2}
\phi_{[k+1]}&=&\phi_{[k]}^{(1)} - \theta_{[k]}^{(1)}\star \theta_{[k]}^{-1}\star \phi_{[k]}
= \left|
\begin{array}{cc}
\theta_{[k]} & \phi_{[k]}\\
\theta_{[k]}^{(1)} & \fbox{$\phi_{[k]}^{(1)}$}
\end{array}
\right|_\star,~~~\phi_{[1]}=\phi,
\\
J_{[k+1]}&=& - \theta_{[k]}^{(1)}\star \theta_{[k]}^{-1}\star J_{[k]}
= \left|
\begin{array}{cc}
\theta_{[k]} & 1\\
\theta_{[k]}^{(1)} & \fbox{$0$}
\end{array}
\right|_\star\star J_{[k]},~~~J_{[1]}=J,
\end{eqnarray}
where $\theta_{[k]}=\phi_{[k]}\vert_{\phi\rightarrow \theta_k}$. 
As is commented at the end of section 3, 
the quasideterminants are well-defined in this situation
{with} all elements $\theta_{[k]}, \phi_{[k]},
\theta_{[k]}^{(1)}, \phi_{[k]}^{(1)}$ {being} $N\times N$ matrices. 
Then the solution to the linear system \eqref{lin_yang} is
\begin{eqnarray}
\label{phi_n+1}
\phi_{[n+1]}
=
\left|
  \begin{array}{cc}
   \Theta& \phi\\
   \vdots &  \vdots\\
   \Theta^{(n-1)}& \phi^{(n-1)}\\
   \Theta^{(n)}& \fbox{$\phi^{(n)}$}
\end{array}\right|_\star,~ 
\end{eqnarray}
where $\Theta$ depends on 
how many times the Darboux transformation is iterated.
It is defined by the $N\times iN$ matrix 
$ \Theta:=(\theta_1,\cdots, \theta_i)$ for $i$ iterations.
The simbol $\Theta^{(k)}$ is defined by $\Theta^{(k)}:= \Theta \Lambda^k$ and 
$\Lambda:=\mbox{diag}(\Lambda_1,\cdots,\Lambda_i)$. Note that in \eqref{phi_n+1}, $\Theta^{(k)}$ is {an} $N\times nN$ matrix with  $\Theta=(\theta_1,\cdots, \theta_n)$ and $\Lambda=\mbox{diag}(\Lambda_1,\cdots,\Lambda_n)$.

\vspace{3mm}
\noindent
(Proof) Proof is made by induction. It holds for $n=1$ by definition.
Assuming that it holds for some $k~ (1 \leq k \leq n)$, 
we consider
\begin{eqnarray*}
Q:= \phi_{[k+2]}=
\phi_{[k+1]}^{(1)} - \theta_{[k+1]}^{(1)}\star 
\theta_{[k+1]}^{-1}\star \phi_{[k+1]}
= \left|
\begin{array}{cc}
\theta_{[k+1]} & \phi_{[k+1]}\\
\theta_{[k+1]}^{(1)} & \fbox{$\phi_{[k+1]}^{(1)}$}
\end{array}
\right|_\star.
\end{eqnarray*}

The first term in $Q$ is
\begin{eqnarray*}
 \phi_{[k+1]}^{(1)}=
 \phi_{[k+1]} \zeta=
\left|
  \begin{array}{cc}
   \Theta&\phi\zeta\\
   \vdots &  \vdots\\
   \Theta^{(k-1)}& \phi^{(k-1)}\zeta\\
   \Theta^{(k)}&  \fbox{$\phi^{(k)}\zeta$}
\end{array}\right|_\star=
\left|
  \begin{array}{cc}
   \Theta\Lambda& \phi\zeta\\
   \vdots &   \vdots\\
   \Theta^{(k-1)}\Lambda& \phi^{(k-1)}\zeta\\
   \Theta^{(k)}\Lambda&  \fbox{$\phi^{(k)}\zeta$}
\end{array}\right|_\star=
\left|
  \begin{array}{cc}
   \Theta^{(1)}& \phi^{(1)}\\
   \vdots &   \vdots\\
   \Theta^{(k)}& \phi^{(k)}\\
   \Theta^{(k+1)}&  \fbox{$\phi^{(k+1)}$}
\end{array}\right|_\star,
\end{eqnarray*}
where we have applied proposition 3.1. 
Note that $\Theta=(\theta_1,\cdots,\theta_k)$ here. 
Next by using homological relation, 
\begin{eqnarray}
\theta_{[k+1]}^{-1}\star \phi_{[k+1]}\nonumber
&=&
\left|
  \begin{array}{cc}
   \Theta& \theta_{k+1}\\
   \vdots &  \vdots\\
   \Theta^{(k-1)}& \theta_{k+1}^{(k-1)}\\
   \Theta^{(k)}&  \fbox{$\theta_{k+1}^{(k)}$}
\end{array}\right|_\star^{-1}\star
\left|
  \begin{array}{cc}
   \Theta& \phi\\
   \vdots &  \vdots\\
   \Theta^{(k-1)}& \phi^{(k-1)}\\
   \Theta^{(k)}&  \fbox{$\phi^{(k)}$}
\end{array}\right|_\star\\\nonumber
&=&
\left(
\left|
  \begin{array}{cc}
   \Theta& 1\\
   \Theta^{(1)}   & 0 \\
   \vdots &   \vdots\\
   \Theta^{(k-1)}& 0\\
   \Theta^{(k)}&  \fbox{$0$}
\end{array}\right|_\star\star
\left|
  \begin{array}{cc}
   \Theta& \fbox{$\theta_{k+1}$}\\
   \vdots   & \vdots\\
   \Theta^{(k-1)}& \theta_{k+1}^{(k-1)}\\
   \Theta^{(k)}&  \theta_{k+1}^{(k))}
\end{array}\right|_\star
\right)^{-1}\star
\left|
  \begin{array}{cc}
   \Theta& \phi\\
   \vdots   & \vdots\\
   \Theta^{(k-1)}& \phi^{(k-1)}\\
   \Theta^{(k)}&  \fbox{$\phi^{(k)}$}
\end{array}\right|_\star\\\nonumber
&=&
\left|
  \begin{array}{cc}
   \Theta&\fbox{$\theta_{k+1}$}\\
   \vdots &   \vdots\\
   \Theta^{(k-1)}& \theta_{k+1}^{(k-1)}\\
   \Theta^{(k)}&  \theta_{k+1}^{(k)}
\end{array}\right|_\star^{-1}\star
\left|
  \begin{array}{cc}
   \Theta& \fbox{$\phi$}\\
   \vdots &   \vdots\\
   \Theta^{(k-1)}& \phi^{(k-1)}\\
   \Theta^{(k)}&  \phi^{(k)}
\end{array}\right|_\star. 
\end{eqnarray}
It then follows immediately by using NC Jacobi identity
that 
\begin{eqnarray*}
Q=
\left|
  \begin{array}{ccc}
   \Theta&\theta_{k+1}& \phi\\
   \Theta^{(1)}  &\theta_{k+1}^{(1)} &  \phi^{(1)} \\
   \vdots &  \vdots & \vdots\\
   \Theta^{(k)}&\theta_{k+1}^{(k)}& \phi^{(k)}\\
   \Theta^{(k+1)} &\theta_{k+1}^{(k+1)}& \fbox{$\phi^{(k+1)}$}
\end{array}\right|_{\star .} \square
\end{eqnarray*}

\vspace{3mm}
\noindent
{\bf Theorem 4.3}
Yang's $J$ matrix generated from a trivial seed solution $J=1$
is represented by a single quasideterminant as follows:
\begin{eqnarray}
J_{[n+1]}=
\left|
  \begin{array}{cc}
   \Theta& 1\\
   \Theta^{(1)} & 0\\
   \vdots   & \vdots\\
   \Theta^{(n-1)}& 0\\
   \Theta^{(n)}&  \fbox{$0$}
\end{array}\right|_\star,
\end{eqnarray}

\noindent
(Proof) This follows immediately by induction and
the definition \eqref{Darboux_J} and the formula
\begin{eqnarray*}
-\theta_{[k+1]}^{(1)}\theta_{[k+1]}^{-1}
=
\left|
  \begin{array}{ccc}
   \Theta&\theta_{k+1} & 1\\
   \Theta^{(1)}&\theta_{k+1}^{(1)} & 0\\
   \vdots &  ~~~  & \vdots\\
   \Theta^{(k)}&\theta_{k+1}^{(k)}& 0\\
   \Theta^{(k+1)}&\theta_{k+1}^{(k+1)} & \fbox{$0$}
\end{array}\right|_\star 
\star
\left|
  \begin{array}{cc}
   \Theta& 1\\
   \Theta^{(1)}& 0\\
   \vdots &   \vdots\\
   \Theta^{(k-1)}& 0\\
   \Theta^{(k)}&  \fbox{$0$}
\end{array}\right|_{\star .}^{-1}  \square
\end{eqnarray*}


\vspace{3mm}
\noindent
{\bf Theorem 4.4}
Yang's $K$ matrix generated from a trivial seed solution $K=0$ 
is represented by a single quasideterminant as follows:
\begin{eqnarray}
K_{[n+1]}=
-\left|
  \begin{array}{cc}
   \Theta& 0\\
   \vdots &   \vdots\\
   \Theta^{(n-2)}& 0\\
   \Theta^{(n-1)}& 1\\
   \Theta^{(n)}&  \fbox{$0$}
\end{array}\right|_\star.
\end{eqnarray}

\noindent
(Proof) First we note that the recursion relation on $\Theta$ 
holds by definition:
\begin{eqnarray}
\label{recursion}
 \partial_y\Theta-\partial_{\overline{z}}\Theta \cdot\Lambda=0,~~~
 \partial_z\Theta+\partial_{\overline{y}}\Theta \cdot\Lambda=0.
\end{eqnarray}

By using the derivative formula \eqref{row_diff}
and the recursion relation \eqref{recursion}, we have
\begin{eqnarray*}
\partial_y J_{[k+1]}\star J_{[k+1]}^{-1}\nonumber
\!\!\!&=&\!\!\!
\left(
\left|
  \begin{array}{cc}
   \Theta& 1\\
   \Theta^{(1)} & 0\\
   \vdots  & \vdots\\
   \Theta^{(k-1)}& 0\\
   \partial_y\Theta^{(k)}& \fbox{$0$}
\end{array}\right|_\star
\!\!+\sum_{p=0}^{k-1}
\left|
  \begin{array}{cc}
   \Theta& 0\\
   \vdots &  \vdots\\
   \Theta^{(p)}& 1\\
   \vdots  & \vdots\\
   \Theta^{(k-1)}& 0\\
   \Theta^{(k)} & \fbox{$0$}
\end{array}\right|_\star \star
\left|
  \begin{array}{cc}
   \Theta&1\\
   \Theta^{(1)} & 0\\
   \vdots & \vdots\\
   \Theta^{(k-1)}& 0\\
   \partial_y\Theta^{(k)} & \fbox{$0$}
\end{array}\right|_\star
\right)
\star
\left|
  \begin{array}{cc}
   \Theta& 1\\
   \Theta^{(1)} & 0\\
   \vdots  & \vdots\\
   \Theta^{(k-1)}& 0\\
   \Theta^{(k)} & \fbox{$0$}
\end{array}\right|_\star^{-1}\\\nonumber
\!\!\!&=&\!\!\!\!
\left(
\left|
  \begin{array}{cc}
   \Theta& 1\\
   \Theta^{(1)} & 0\\
   \vdots  & \vdots\\
   \Theta^{(k-1)}& 0\\
   \partial_{\overline{z}}\Theta^{(k+1)} & \fbox{$0$}
\end{array}\right|_\star
\!\!+\sum_{p=0}^{k-1}
\left|
  \begin{array}{cc}
   \Theta&0\\
   \vdots & \vdots\\
   \Theta^{(p)} & 1\\
   \vdots  & \vdots\\
   \Theta^{(k-1)}& 0\\
   \Theta^{(k)}& \fbox{$0$}
\end{array}\right|_\star \!\!\star \!
\left|
  \begin{array}{cc}
   \Theta&1\\
   \Theta^{(1)} & 0\\
   \vdots &   \vdots\\
   \Theta^{(k-1)}& 0\\
   \partial_{\overline{z}}\Theta^{(k+1)} & \fbox{$0$}
\end{array}\right|_\star
\right)\!
\star
\left|
  \begin{array}{cc}
   \Theta& 1\\
   \Theta^{(1)} & 0\\
   \vdots  & \vdots\\
   \Theta^{(k-1)}& 0\\
   \Theta^{(k)} & \fbox{$0$}
\end{array}\right|_\star^{-1}\\\nonumber
\!\!\!&=&\!\!\!
-\left|
  \begin{array}{cc}
   \Theta&0\\
   \Theta^{(1)} & 0\\
   \vdots  & \vdots\\
   \Theta^{(k-1)}& 1\\
   \partial_{\overline{z}}\Theta^{(k)} & \fbox{$0$}
\end{array}\right|_\star
\!\!-\sum_{p=0}^{k-1}
\left|
  \begin{array}{cc}
   \Theta&0\\
   \vdots  & \vdots\\
   \Theta^{(p)} & 1\\
   \vdots  & \vdots\\
   \Theta^{(k-1)}& 0\\
   \Theta^{(k)} & \fbox{$0$}
\end{array}\right|_\star \!\star
\left|
  \begin{array}{cc}
   \Theta& 0\\
   \Theta^{(1)} & 0\\
   \vdots   & \vdots\\
   \Theta^{(k-1)}& 1\\
   \partial_{\overline{z}}\Theta^{(k)} & \fbox{$0$}
\end{array}\right|_\star
\!\!=\!-\partial_{\overline{z}}\left|
  \begin{array}{cc}
   \Theta& 0\\
   \Theta^{(1)} & 0\\
   \vdots   & \vdots\\
   \Theta^{(k-1)}&1\\
   \Theta^{(k)}& \fbox{$0$}
\end{array}\right|_\star
\end{eqnarray*}
where the final line uses
the following formula, 
obtained by mimicking part of the proof of 
the derivative formula \eqref{row_diff}
\begin{eqnarray*}
\left|
  \begin{array}{cc}
   \Theta&1\\
   \Theta^{(1)} & 0\\
   \vdots &   \vdots\\
   \Theta^{(k-1)}& 0\\
   \partial_{\overline{z}}\Theta^{(k+1)} & \fbox{$0$}
\end{array}\right|_\star
\!\!\!&=&\!\!\!
-\partial_{\overline{z}}\Theta\Lambda^{k+1}
\star
\left(
\begin{array}{c}
\Theta \\
\Theta^{(1)} \\
\vdots \\
\Theta^{(k-1)}
\end{array}
\right)^{-1}
\left(
\begin{array}{c}
1 \\
0 \\
\vdots \\
0
\end{array}
\right)\\
\!\!\!&=&\!\!\!
-\partial_{\overline{z}}\Theta\Lambda^{k}
\star
\left(
\begin{array}{c}
\Theta \\
\Theta^{(1)} \\
\vdots \\
\Theta^{(k-1)}
\end{array}
\right)^{-1}\!\!\!\!
\star
\sum_{p=1}^{k}e_p e_p^t
\left(
\begin{array}{c}
\Theta \\
\Theta^{(1)} \\
\vdots \\
\Theta^{(k-1)}
\end{array}
\right)
\Lambda\star
\left(
\begin{array}{c}
\Theta \\
\Theta^{(1)} \\
\vdots \\
\Theta^{(k-1)}
\end{array}
\right)^{-1}
\left(
\begin{array}{c}
1 \\
0 \\
\vdots \\
0
\end{array}
\right)\\
\!\!\!&=&\!\!\!
-\sum_{p=0}^{k-1}
\left|
  \begin{array}{cc}
   \Theta&0\\
   \vdots  & \vdots\\
   \Theta^{(p)} & 1\\
   \vdots  & \vdots\\
   \Theta^{(k-1)}& 0\\
   \partial_{\overline{z}}\Theta^{(k)} & \fbox{$0$}
\end{array}\right|_\star \!\! \star \!
\left|
  \begin{array}{cc}
   \Theta& 1\\
   \Theta^{(1)} & 0\\
   \vdots   & \vdots\\
   \Theta^{(k-1)}& 0\\
   \Theta^{(p+1)}& \fbox{$0$}
\end{array}\right|_\star
\!\!=\!-
\left|
  \begin{array}{cc}
   \Theta&0\\
   \Theta^{(1)} & 0\\
   \vdots  & \vdots\\
   \Theta^{(k-1)}& 1\\
   \partial_{\overline{z}}\Theta^{(k)} & \fbox{$0$}
\end{array}\right|_\star \!\!\star \!
\left|
  \begin{array}{cc}
   \Theta& 1\\
   \Theta^{(1)} & 0\\
   \vdots   & \vdots\\
   \Theta^{(k-1)}& 0\\
   \Theta^{(k)}& \fbox{$0$}
\end{array}\right|_{\star .} 
\end{eqnarray*}

\noindent
Gauge fields are obtained from these potentials: 
\begin{eqnarray}
\label{A}
A_z^{[n+1]}
= -\partial_{z}J_{[n+1]}\star J_{[n+1]}^{-1} = 
-\partial_{\overline{y}} K_{[n+1]},~
A_y^{[n+1]}
= -\partial_{y}J_{[n+1]}
\star J_{[n+1]}^{-1} = \partial_{\overline{z}} K_{[n+1]}.
\end{eqnarray}


\section{Soliton Solutions of NC ASDYM equation}

In this section, we discuss 
exact soliton solutions for $G=GL(N)$
focusing on $N=2$ solutions. 
In order to discuss energy densities of the solutions,
we put a suitable reduction condition 
so that the energy densities can be real-valued.
In subsection 5.1, we discuss one soliton solutions 
and calculate the energy density explicitly. 
We prove that the energy density is real-valued 
and localized on a three-dimensional hyperplane
in $\mathbb{R}^4$. Hence this soliton is a domain-wall 
in the 4-dimensional Euclidean space. 
In subsection 5.2, we study 
asymptotic behavior of 
multi soliton solutions of the same type
where the configurations take real values
in the asymptotic region. 

Let us first introduce star-exponential functions by
\begin{eqnarray*}
e_\star^{f(x)}:=\sum_{n=0}^{\infty}
\frac{1}{n!}\underbrace{f(x)\star \cdots \star f(x)}_{n {\scriptsize
\mbox{times}}}.
\end{eqnarray*}
When $f(x)$ is a linear function of $x^\mu$, 
the star exponential function $e_\star^{f(x)}$ is tractable.
For example, 
\begin{eqnarray}
\label{inverse_exp}
 (e_\star^{k_\mu x^\mu})^{-1} &=& e_\star^{-k_\mu x^\mu},\\
 \partial_{\nu} e_\star^{k_\mu x^\mu}
  &=& k_{\nu} e_\star^{k_\mu x^\mu}.
\label{der_exp}
\end{eqnarray}
These formula play crucial roles in discussion
on behavior of noncommutative soliton solutions.
Soliton solutions for $G=GL(N)$ are given as follows.

\vspace{3mm}
\noindent
{\bf Theorem 5.1}
Soliton-type solutions for $G=GL(N)$ 
can be constructed 
when the $N\times N$ matrices $\Lambda_{s}=$
diag$(\lambda^{(s)}_1,\cdots, \lambda^{(s)}_N)~(s=1,2,\cdots, n)$
and the associated $N\times N$ eigenfunctions are
\begin{eqnarray}
\label{Nsoliton_sol}
 \theta_s&=&\left(
\begin{array}{ccc}
a^{(s)}_{11}e_\star^{K^{(s)}_1}
+a^{\prime (s)}_{11}e_\star^{-K^{(s)}_1}
& \cdots  &
a^{(s)}_{1N}e_\star^{K^{(s)}_N}
+a^{\prime (s)}_{1N}e_\star^{-K^{(s)}_N}
 \\ 
\vdots  & & \vdots \\
a^{(s)}_{N1}e_\star^{K^{(s)}_1}
+a^{\prime (s)}_{N1}e_\star^{-K^{(s)}_1}
  & \cdots & a^{(s)}_{NN}e_\star^{K^{(s)}_N}
+a^{\prime (s)}_{NN}e_\star^{-K^{(s)}_N} 
\end{array}\right),\\
K^{(s)}_p 
&:=&\lambda^{(s)}_p \beta^{(s)}_p z
+\alpha^{(s)}_p\overline{z}
+\lambda^{(s)}_p \alpha^{(s)}_p y
-\beta^{(s)}_p\overline{y}.
\end{eqnarray}
where $a_{pq}^{(s)}, a^{\prime(s)}_{pq}, 
\alpha^{(s)}_p, \beta^{(s)}_p ~(p,q=1,2,\cdots,N)$ 
are complex constants. 
We can easily check that this solves \eqref{recursion}. 

\vspace{3mm}
\noindent
{\bf Corollary 5.2}
There is no non-trivial soliton solution 
\eqref{Nsoliton_sol} for $G=GL(1)$ as in 
theorem 5.1. 

\noindent
(Proof) For $G=GL(1)$, the eigenvalue matrix 
$\Lambda_s$ is a $1\times 1$ scalar function
which is commutative. 
Then the solution of the noncommutative Yang's equation 
becomes constant:
\begin{eqnarray*}
J_{[n+1]}=
\left|
  \begin{array}{cccc}
 \theta_1 \! &\cdots&\!\theta_n \! &\! 1\\
 \theta_1\Lambda_1\!&\cdots&\!\theta_n\Lambda_n \! &\! 0\\
 \vdots \! & & &\!\vdots\\
 \theta_1 \Lambda_1^{n-1}\!&\cdots&\!\theta_n\Lambda_n^{n-1}\!&\! 0\\
 \theta_1 \Lambda_1^{n}\!&\cdots&\!\theta_n\Lambda_n^{n}\!&\!  \fbox{$0$}
\end{array}\right|_\star
\!\!=\!
\left|
  \begin{array}{cccc}
 \theta_1 \!&\cdots&\!\theta_n \! &\! 1\\
 \Lambda_1\theta_1\!&\cdots&\!\Lambda_n\theta_n \!&\! 0\\
 \vdots \!  & & \!&\!\vdots\\
 \Lambda_1^{n-1}\theta_1\! &\cdots&\!\Lambda_n^{n-1}\theta_n\!&\! 0\\
 \Lambda_1^{n}\theta_1\!&\cdots&\!\Lambda_n^{n}\theta_n\!& \!\fbox{$0$}
\end{array}\right|_\star
\!\!=\!
\left|
  \begin{array}{cccc}
 1  \!&\cdots&  \! 1  \!& \! 1\\
 \Lambda_1\!&\cdots&\!\Lambda_n  \! &\! 0\\
 \vdots  \! & & \!&\!\vdots\\
 \Lambda_1^{n-1}\! &\cdots&\!\Lambda_n^{n-1}\!&\! 0\\
 \Lambda_1^{n}\!&\cdots&\!\Lambda_n^{n}\!& \!\fbox{$0$}
\end{array}\right|_\star.
\end{eqnarray*}
In the final step, we eliminate a common factor 
$\theta_s$ on the right side of 
the $s$-th column of the quasideterminant. $\square$


Finally we make a comment on an important formula 
which play crucial roles in the following discussion.
Let $x^\rho,x^\sigma$ be noncommutative 
space-time coordinates $[x^\rho,x^\sigma]_\star=i\vartheta$. 
Introducing new noncommutative coordinates as 
$w:=x^\rho+k x^\sigma, \overline{w}:=x^\rho-kx^\sigma~
(k\in \mathbb{C})$, we can easily find
\begin{eqnarray}
\label{hol}
 f(w)\star g(w)= f(w) g(w)
\end{eqnarray}
because the star-product (\ref{star}) is rewritten in terms of
$(w,\overline{w})$ as 
\begin{eqnarray}
\label{hol}
 f(w,\overline{w})\star g(w,\overline{w})=
e^{ik \vartheta
\left(
\partial_{\overline{w}_1}
\partial_{w_2}-
\partial_{w_1}
\partial_{\overline{w}_2}
\right)}
f(w_1,\overline{w}_1)
g(w_2,\overline{w}_2) \Big{\vert}_{\scriptsize
\begin{array}{c} w_1=w_2=w\\
\overline{w}_1 =\overline{w}_2=\overline{w}. \end{array}}
\label{hol}
\end{eqnarray}
In four-dimensional noncommutative spaces, 
for any functions $f$ which have the forms of $f(k_\mu x^{\mu})$,  
the property \eqref{hol} holds 
for any pair of noncommutative coordinates $x^\rho$
and $x^\sigma$ among the four real coordinates $x^\mu~(\mu=1,2,3,4)$
and therefore the star-products is reduced to ordinary
commutative products of functions.

\subsection{One-Soliton Solutions of NC ASDYM equation}

In this subsection, we focus on one-soliton solutions for 
$J\in U(2)$. 
In commutative spaces,  
they are obtained and energy densities of them are 
real-valued \cite{HaHu}.
A noncommutative candidate is given by:
\begin{eqnarray}
\label{U2_1}
\theta=\left(
\begin{array}{cc}
ae_\star^{K}
& 
be_\star^{-\overline{K}}
 \\ 
-be_\star^{-K}
  & 
ae_\star^{\overline{K}}
\end{array}\right),~~~
\Lambda=
\left(
\begin{array}{cc}
\lambda & 0 \\
0 & -\lambda
\end{array}
\right),~~~\lambda=e^{i\varphi}
\end{eqnarray}
where 
$K:= 
\lambda \beta z
+\alpha\overline{z}
+\lambda \alpha y
-\beta\overline{y}$, $a,b,\varphi
\in \mathbb{R}$, and 
$\alpha,\beta\in \mathbb{C}$. 
In this subsection we prove that this solution
leads to real-valued one-soliton solution 
in every region of $\mathbb{R}^4$. 

Yang's $J$-matrix becomes
\begin{eqnarray}
J&=&-\theta \Lambda \star \theta^{-1}
=
\left|
\begin{array}{cc}
\theta & 1\\
\theta \Lambda & \fbox{$0$}
\end{array}
\right|_\star\nonumber\\
&=&
\left|
\begin{array}{cc}
\begin{array}{cc}
ae_\star^{K}
& 
be_\star^{-\overline{K}}
 \\ 
-be_\star^{-K}
  & 
ae_\star^{\overline{K}}
\end{array}
& 1_{2\times 2}\\
\begin{array}{cc}
\lambda ae_\star^{K}
& 
-\lambda be_\star^{-\overline{K}}
 \\ 
-\lambda be_\star^{-K}
  & 
-\lambda ae_\star^{\overline{K}}
\end{array}
& \fbox{$O_{2\times 2}$}
\end{array}
\right|_\star
=
\left|
\begin{array}{cc}
\begin{array}{cc}
ae_\star^{K+\overline{K}+i\Delta}
& 
be_\star^{-(K+\overline{K})-i\Delta}
 \\ 
-be_\star^{\overline{K}-K-i\Delta}
  & 
ae_\star^{\overline{K}-K+i\Delta}
\end{array}
& 1_{2\times 2}\\
\begin{array}{cc}
\lambda ae_\star^{K+\overline{K}+i\Delta}
& 
-\lambda be_\star^{-(K+\overline{K})-i\Delta}
 \\ 
-\lambda be_\star^{\overline{K}-K-i\Delta}
  & 
-\lambda ae_\star^{\overline{K}-K+i\Delta}
\end{array}
& \fbox{$O_{2\times 2}$}
\end{array}
\right|_\star, \nonumber\\
&&
{\mbox{where }} \Delta:=
(1/2)k_\mu \overline{k}_\nu
\vartheta^{\mu\nu},~~K=k_\mu x^\mu. 
\label{k_mu}
\end{eqnarray}
In the final step, we use invariance of quasideterminant
by right-multiplication of 
$e_\star^{\overline{K}}$ on the first column 
and 
$e_\star^{-K}$ on the second column, 
and use the formula $e_\star^{K}\star 
e_\star^{\overline{K}}=e_\star^{K+\overline{K}+i\Delta}$. 

Here let us take the following 
gauge transformation on $\theta$:
\begin{eqnarray*}
 \theta \mapsto g^{-1}\star \theta,~~~
g^{-1}=\left(
\begin{array}{cc}
1&0 \\
0&e_\star^{K-\overline{K}}. 
\end{array}
\right),~~~g\star g^\dagger=g^\dagger\star g=1
\end{eqnarray*}
Then we get 
\begin{eqnarray}
J=
\left|
\begin{array}{cc}
\begin{array}{cc}
ae_\star^{K+\overline{K}+i\Delta}
& 
be_\star^{-(K+\overline{K})-i\Delta}
 \\ 
-be^{-i\Delta}
  & 
ae^{i\Delta}
\end{array}
& 1_{2\times 2}\\
\begin{array}{cc}
\lambda ae_\star^{K+\overline{K}+i\Delta}
& 
-\lambda b e_\star^{-(K+\overline{K})-i\Delta}
 \\ 
-\lambda b e^{-i\Delta}
  & 
-\lambda ae^{i\Delta}
\end{array}
& \fbox{$O_{2\times 2}$}
\end{array}
\right|_\star.
\label{J_final}
\end{eqnarray}
Now we can see that the coordinate 
dependence has the form of $K+\overline{K}
=(k_\mu+\overline{k}_\mu)x^\mu$
and the star-products disappear after this step 
due to \eqref{hol}.
In order to get real-valued energy density,
$J$ should be unitary, which is equivalent 
that in the final form \eqref{J_final} of $J$, 
$ae^{i\Delta}$ and $be^{-i\Delta}$ should be real.
This is realized by redefinition of $a\mapsto a e^{-i\Delta}$ 
and $b\mapsto b e^{i\Delta}$. 

Results are summarized in the following:

\vspace{3mm}
\noindent
{\bf Theorem 5.3}
One soliton solution 
for $J\in U(2)$ is given by 
\begin{eqnarray}
\label{U2_1_summary}
\theta=\left(
\begin{array}{cc}
ae_\star^{K-i\Delta}
& 
be_\star^{-(\overline{K}-i\Delta)}
 \\ 
-be_\star^{-(K-i\Delta)}
  & 
ae_\star^{\overline{K}-i\Delta}
\end{array}\right)
,~~~
\Lambda=
\left(
\begin{array}{cc}
e^{i\varphi} & 0 \\
0 & -e^{i\varphi}
\end{array}
\right),~~~a,b, \varphi\in \mathbb{R}
\end{eqnarray}
where
$K
=e^{i\varphi} \beta z
+\alpha\overline{z}
+e^{i\varphi} \alpha y
-\beta\overline{y}$, $\alpha,\beta\in \mathbb{C}$. 

This solution $\theta$ leads to 
the final form \eqref{J_final} of $J$
without the constant $\Delta$. 
Gauge fields are calculated as in \eqref{A} 
from this $J$
and field strengths are obrtained. 
In the calculations, the star-products always 
disappear due to the property \eqref{hol} and 
the energy density is reduced to the commutative one \cite{HaHu}:
\begin{eqnarray}
{\mbox{Tr}} F_\star^2= 
{\mbox{Tr}} F^2= 
32(\vert\alpha\vert^2+\vert \beta\vert^2)
\left(
2{\mbox{sech}}^2 X 
-3
{\mbox{sech}}^4 X 
\right), 
\end{eqnarray}
where 
$X:=K+ \overline{K}+\log \vert a/b\,\vert$.
This is actually real-valued at every point in $\mathbb{R}^4$. 
We can find that the energy density has its peak on 
a three-dimensional hyperplane
defined by $X=K+ \overline{K}+\log \vert a/b\,\vert=0$
whose normal vector is $k_\mu+\overline{k}_\mu$. 
($k_\mu$ is defined in \eqref{k_mu}.)  
This is a domain-wall in $\mathbb{R}^4$. 

\subsection{Multi-Soliton Solutions of NC ASDYM equation}

In this subsection, 
we present multi-soliton solutions of the same type
as the previous subsection 
and discuss the asymptotic behavior of them. 

The $n$-soliton solution is given by ($s=1,2,\cdots,n$):
\begin{eqnarray}
\label{U2_n}
\theta_s=\left(
\begin{array}{cc}
a_s e_\star^{K_s-i\Delta_s}
& 
b_s e_\star^{-(\overline{K_s}-i\Delta_s)}
 \\ 
-b_s e^{-(K_s-i\Delta_s)}
  & 
a_s e^{\overline{K_s}-i\Delta_s}
\end{array}\right)
,~~~
\Lambda_s=
\left(
\begin{array}{cc}
e^{i\varphi_s} & 0 \\
0 & -e^{i\varphi_s}
\end{array}
\right),
\end{eqnarray}
where
$K_s
=e^{i
\varphi_s} \beta_s z
+\alpha_s\overline{z}
+e^{i\varphi_s} \alpha_s y
-\beta_s\overline{y}=:k_\mu^{(s)} x^\mu$, 
$\Delta_s:=(1/2)k^{(s)}_\mu \overline{k}_{\nu}^{(s)}
\vartheta^{\mu\nu}
$, 
and 
$\alpha_s,\beta_s\in \mathbb{C},~
a_s, b_s, \varphi_s\in \mathbb{R}$. 
We assume that $K_s$ $(s=1,2,\cdots,n)$ are independent
and there is no special relation among them,
which corresponds to pure-soliton scattering, 
that is, no resonance.

As in the standard discussion, let us consider
the asymptotic limit $r:=((x^1)^2+(x^2)^2+(x^3)^2+(x^4)^2)^{1/2}
\rightarrow \infty$ keeping both $K_I$ and $\overline{K_I}$ finite.
(Later this condition will be loosed.)  
Then for $s\neq I$, $\vert e_\star^{K_s} \vert $ and 
$\vert e_\star^{\overline{K_s}}\vert $ 
go to positive infinity or zero. 
Then we can get the asymptotic form of $J$ by eliminating 
common factors which go to infinite in a column of 
the quasideterminant $J$: 
\begin{eqnarray*}
J \rightarrow 
\left|
\begin{array}{cccccc}
C_1
&
\cdots
&
\theta_I
& 
\cdots
&
C_n
& 
1
\\
C_1 \Lambda_1
&
\cdots
&
\theta_I\Lambda_I
&
\cdots
& 
C_n\Lambda_n
&0
\\
\vdots
&
&
\vdots
&
& 
\vdots
&
\vdots
\\
C_1\Lambda_1^n
& 
\cdots
&
\theta_I\Lambda_1^n
& 
\cdots
&
C_n\Lambda_n^n
& \fbox{0}
\end{array}
\right|_\star
~~~
\mbox{where}
~
 C_s=\left(
\begin{array}{cc}
1&0 \\
0&1 
\end{array}
\right)
~\mbox{or}~
\left(
\begin{array}{cc}
0&1 \\
-1&0 
\end{array}
\right).
\end{eqnarray*}
Focusing on 
the $\theta_I$-column in the above representation of $J$,
let us multiply 
$e_\star^{\overline{K}_I}$ on the first column of $\theta_I$
and 
$e_\star^{-K_I}$ on the second column of $\theta_I$, 
and further take a gauge transformation
$\Theta \mapsto g_I^{-1}\star \Theta,~
g_I^{-1}={\mbox{diag}}(1,e_\star^{K_I-\overline{K}_I}),~
\Theta=(\theta_1,\cdots,\theta_n)$
as in subsection 5.1. Then we get 
\begin{eqnarray*}
J \rightarrow 
\left|
\begin{array}{cccccc}
C_1
&
\cdots
&
\widetilde{\theta}_I
& 
\cdots
&
C_n
& 
1
\\
C_1 \Lambda_1
&
\cdots
&
\widetilde{\theta}_I\Lambda_I
&
\cdots
& 
C_n\Lambda_n
&0
\\
\vdots
&
&
\vdots
&
& 
\vdots
&
\vdots
\\
C_1\Lambda_1^n
& 
\cdots
&
\widetilde{\theta}_I
\Lambda_1^n
& 
\cdots
&
C_n\Lambda_n^n
& \fbox{0}
\end{array}
\right|_\star
~~~
\widetilde{\theta}_I=\left(
\begin{array}{cc}
a_{I}
 e_\star^{K_{I}+\overline{K_{I}}}
& 
b_{I} e_\star^{-(K_{I}+\overline{K_{I}})}
 \\ 
-b_{I} 
  & 
a_{I} 
\end{array}\right).
\end{eqnarray*}
We note that the factors $e_\star^{K_I-\overline{K}_I}$
in $C_1,\cdots, C_n$ disappear because it is 
a common factor in a column of the quasideterminant.
Now we can see that the coordinate 
dependence has the form of $K_I+\overline{K}_I\in \mathbb{R}$
and the star-products disappear after this step.
All elements are real now and the energy density should be real. 

Hence we can conclude that in the asymptotic behavior
is all the same as commutative case in 
the pure soliton process. As a result, 
the $n$ soliton solutions would have 
$n$ isolated localized lumps of energy.
They preserve their shapes and velocities of
the localized solitary wave lumps. 
It is worthwhile to compare commutative limits of our solutions
with commutative results \cite{deVega}. 
Details will be reported elsewhere.

\section{Conclusion and Discussion}

In this paper, we gave multi soliton solutions
of noncommutative anti-self-dual Yang-Mills equations 
for $G=GL(N)$ by the Darboux transformations 
and discussed soliton solutions for $G=GL(2)$ in detail. 
The generated solutions were represented  
in terms of quasideterminants in compact forms. 
The Yang $J$-matrices 
obtained here are different from 
those obtained by the Penrose-Ward transform \cite{GHN, GHN2}. 
$K$ matrices are new. 
We found that special 
one-soliton solutions for $G=GL(2)$ have 
the same configuration as commutative one which have 
localized energy density on a three-dimensional hyperplane. 
This is a domain-wall in $\mathbb{R}^4$. 
We also presented multi-soliton solutions
of the same type and discussed asymptotic behaviors of them.

The present discussion is straightforwardly 
generalized to other signature such as 
ultra-hyperbolic signature $(+\!+\!--)$. 
The anti-self-dual Yang-Mills equations
with the ultra-hyperbolic signature 
can be reduced to various lower-dimensional 
integrable systems such as KdV, NLS equations.
This is first conjectured in \cite{Ward} (known as 
Ward's conjecture) and summarized in \cite{MaWo}. 
Various examples of the noncommutative Ward conjecture
are seen in \cite{Hamanaka_NPB}. 
This Darboux technique 
could be applied to 
wide class of noncommutative integrable systems.

In commutative spaces, a binary Darboux transformation
is developed in \cite{GNO00}. This gives 
Grammian-type solutions. Noncommutative extension
of it would be possible and interesting (cf. \cite{HaHa}). 
Noncommutative binary Darboux transformations 
are successful for KP and KdV equations \cite{GiNi07} and 
modified KP equations \cite{GNS}. 
Noncommutative extension of the bilinear approach to 
anti-self-dual Yang-Mills equations \cite{MOS} is also 
worthwhile.

In this paper, we discussed pure soliton scatterings. 
It is an interesting question whether there are
resonance processes for (noncommutative) anti-self-dual
Yang-Mills equation with respect to the energy density. 
It might lead to higher-dimensional extension of
Kodama's Grassmannian approach to KP soliton scatterings 
including all possible resonance processes \cite{Kodama}.

It would be also worth clarifying the relation 
between the Wronskian-type solutions in the present paper
and the non-Wronskian-type solutions in \cite{GHN,GHN2}. 
Wronskian-type solutions play crucial roles in 
the Sato theory of the KP equations 
which reveals hidden infinite symmetry on  
infinite-dimensional Grassmannians from the viewpoint 
of Pl\"ucker relations of the Wronskians  \cite{Sato}. 
Hence the present formulation might be a hint for 
a higher-dimensional extension of Sato theory. 
On the other hand, the non-Wronskian solutions relate to
the twistor theory \cite{MaWo, WaWe}. 
The geometrical viewpoints give 
explanation of the origin of integrability and 
can be applied to construction of soliton solutions as well.  
The present Wronskian solutions could be represented  
by the non-Wronskian-type solutions with a suitable parametrization.\footnote{
We note that a simple parameter choice of one soliton solutions
in the non-Wronskian type  
yields trivial energy densities (See section 4 in \cite{HaHu})
while the present Wronskian-type solitons are non-trivial.}
This might shed light on a profound relationship  
between  Sato theory and twistor theory.


\subsection*{Acknowledgments}

The authors would like to thank the organizers at the workshop on 
Integrable systems, special functions and combinatorics (ISLAND5),
Sabhal Mor Ostaig, the Gaelic College, the Isle of Skye, 
23 - 28 June 2019.
MH is also grateful to K.~Takasaki for discussion.
The work of MH is supported 
by Grant-in-Aid for Scientific Research (\#16K05318).
The work of SCH is supported 
by the scholarship of Japan-Taiwan Exchange Association.

\end{document}